Article

# Tunable quantum criticalities in an isospin extended Hubbard model simulator

Qiao Li[1], Bin Cheng[2✉], Moyu Chen[1], Bo Xie[3], Yongqin Xie[1], Pengfei Wang[1], Fanqiang Chen[1], Zenglin Liu[1], Kenji Watanabe[4], Takashi Taniguchi[5], Shi-Jun Liang[1], Da Wang[1], Chenjie Wang[6], Qiang-Hua Wang[1], Jianpeng Liu[3] & Feng Miao[1✉]

Studying strong electron correlations has been an essential driving force for pushing the frontiers of condensed matter physics. In particular, in the vicinity of correlation-driven quantum phase transitions (QPTs), quantum critical fluctuations of multiple degrees of freedom facilitate exotic many-body states and quantum critical behaviours beyond Landau's framework[1]. Recently, moiré heterostructures of van der Waals materials have been demonstrated as highly tunable quantum platforms for exploring fascinating, strongly correlated quantum physics[2–22]. Here we report the observation of tunable quantum criticalities in an experimental simulator of the extended Hubbard model with spin–valley isospins arising in chiral-stacked twisted double bilayer graphene (cTDBG). Scaling analysis shows a quantum two-stage criticality manifesting two distinct quantum critical points as the generalized Wigner crystal transits to a Fermi liquid by varying the displacement field, suggesting the emergence of a critical intermediate phase. The quantum two-stage criticality evolves into a quantum pseudo criticality as a high parallel magnetic field is applied. In such a pseudo criticality, we find that the quantum critical scaling is only valid above a critical temperature, indicating a weak first-order QPT therein. Our results demonstrate a highly tunable solid-state simulator with intricate interplay of multiple degrees of freedom for exploring exotic quantum critical states and behaviours.

Studies of electronic-correlation-driven QPTs have shown many exotic quantum many-body phenomena. Especially, when multiple degrees of freedom are involved, competition of different order parameters and their quantum fluctuations becomes prominent and may lead to new types of quantum critical phases and behaviours beyond Landau's framework. Exploring distinct types of QPTs and their evolution in solid-state platforms would provide unprecedented opportunities to give insight into the origins of non-Landau quantum criticalities. Despite constant attempts, such a platform remains yet to be realized, owing to the lack of capability of in situ and simultaneous tuning of electron correlations and multiple degrees of freedom.

In this work, we demonstrate a new solid-state simulator for the extended Hubbard model residing in cTDBG. Through electrical transport measurements, we demonstrate a generalized Wigner crystal at the filling of $7\frac{2}{3}n_0$, in which the electron correlation can be in situ tailored by varying the perpendicular displacement field. Taking advantage of the decoupled valley and spin degrees of freedom, which give rise to four valley–spin isospin flavours, we—for the first time to our knowledge—observe quantum pseudo and two-stage criticalities and realize in situ evolution of those unconventional quantum criticalities by using a parallel magnetic field.

## Transport signatures of the Wigner crystal state

Our moiré graphene sample was fabricated by stacking two pieces of Bernal-stacked bilayer graphene with opposite chirality; that is, the top bilayer graphene is AB-stacked, whereas the bottom bilayer graphene is BA-stacked, as shown schematically in Fig. 1a. This cTDBG has a twist angle of 0.75° (see Methods), forming moiré patterns consisting of three distinct regions of stacking order ABBA, ABCB and ABAC, as indicated by circles of different colours in Fig. 1b,c. Two graphite gates with applied voltages $V_{tg}$ and $V_{bg}$ are adopted to independently control the carrier density $n$ and displacement field $D$ through $n = C_{tg}V_{tg} + C_{bg}V_{bg}$ and $D = (C_{tg}V_{tg} - C_{bg}V_{bg})/2$, in which $C_{tg}$ ($C_{bg}$) represents the capacitance between cTDBG and the top (bottom) gate.

We first perform four-terminal transport measurements in a perpendicular magnetic field ($B_\perp$) of 200 mT. Figure 1d shows the longitudinal and Hall resistances $R$ and $R_H$ as functions of $n$ measured at displacement field $D = 0$ V nm$^{-1}$ and temperature $T = 1.5$ K. $R$ shows several distinct peaks, corresponding to the charge neutrality point gap at $n = 0$ and the moiré gaps with fourfold spin and valley degeneracy at $n = \pm4n_0$ and $\pm8n_0$, with $n_0$ being the carrier density corresponding to one electron per moiré unit cell. When setting $B_\perp$ to zero and continuously changing $D$, we obtain the 2D map of $R$ as a function of $n$ and $D$ (Fig. 1e), showing tunable resistance at the charge neutrality point and moiré gaps, which is consistent with the displacement-field-dependent band structure in cTDBG (Fig. 1f). Notably, well-defined $R$ peaks develop within the second moiré band on the electron side ($4n_0 < n < 8n_0$) at large $|D|$ (Fig. 2a). Line plots of $R$ at several different $D$ values are shown in Fig. 2b, with marked $R$ peaks gradually appearing at $n = 7n_0$ and $7\frac{2}{3}n_0$. Notably, the $7\frac{2}{3}n_0$ fractional state means that the moiré unit cell is tripled and the supercell accommodates one hole in the second moiré band.

[1]National Laboratory of Solid State Microstructures, School of Physics, Institute of Brain-Inspired Intelligence, Collaborative Innovation Center of Advanced Microstructures, Nanjing University, Nanjing, China. [2]Institute of Interdisciplinary Physical Sciences, School of Science, Nanjing University of Science and Technology, Nanjing, China. [3]School of Physical Science and Technology, ShanghaiTech Laboratory for Topological Physics, ShanghaiTech University, Shanghai, China. [4]Research Center for Functional Materials, National Institute for Materials Science, Tsukuba, Japan. [5]International Center for Materials Nanoarchitectonics, National Institute for Materials Science, Tsukuba, Japan. [6]Department of Physics and HKU-UCAS Joint Institute for Theoretical and Computational Physics at Hong Kong, The University of Hong Kong, Hong Kong, China. ✉e-mail: bincheng@njust.edu.cn; miao@nju.edu.cn

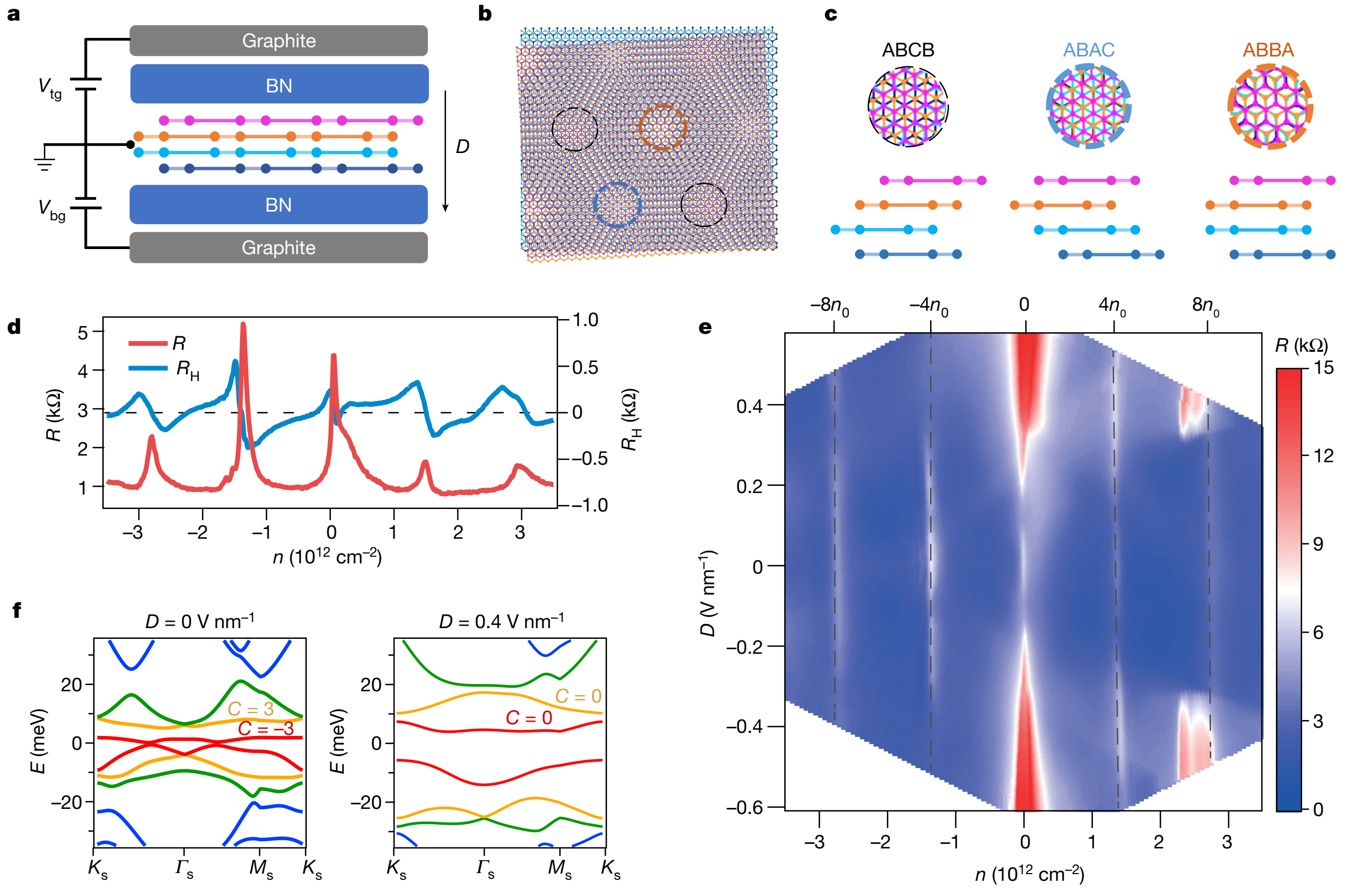

**Fig. 1 | cTDBG with $\theta$ = 0.75°. a**, Schematic of our cTDBG device. Bottom and top gate $V_{bg}$ and $V_{tg}$ are adopted to independently control carrier density $n$ and displacement field $D$. **b**, Moiré patterns of small-twist-angle cTDBG. Three regions of different stacking orders are indicated by circles of different colours. **c**, Top and side views of the different stacked regions marked in **b**. **d**, Longitudinal resistance ($R$) and Hall resistance ($R_H$) as functions of $n$ measured at $D$ = 0 V nm$^{-1}$, $B_\perp$ = 200 mT and $T$ = 1.5 K. **e**, 2D map of $R$ as a function of $n$ and $D$. **f**, Calculated band structures for different displacement fields $D$. The Chern numbers are labelled for the first and second conduction bands.

We refer to this fractional state as a generalized isospin Wigner crystal pinned on periodic moiré potential[23–27], with detailed experimental evidences shown below. First, we obtain the differential resistance d$V$/d$I$ versus d.c. current $I_{ds}$ curves at several different $D$. As shown in Fig. 2c, a notable peak emerges at zero $I_{ds}$ for $|D| \geq 0.33$ V nm$^{-1}$, corresponding to strong non-linearity in the $I$–$V$ curve. In the quantum Wigner crystal state, electrons are localized and pinned by the moiré superlattices, and depinning those electrons requires a finite external voltage, which gives rise to a highly non-linear $I$–$V$ characteristic[28,29]. Increasing the bias current or decreasing $|D|$ will drive depinning of the electrons, making the Wigner crystal finally melt into a metal that exhibits a linear $I$–$V$ characteristic. Similarly, when the temperature increases, the Wigner crystal melts by means of thermal fluctuations and, thus, the d$V$/d$I$ peak gradually fades until it totally disappears (Fig. 2d). Second, we investigate the thermal activation behaviour of this fractional filling state, as presented in Fig. 2e. We set $D$ at −0.46 V nm$^{-1}$, which is within the insulating regime, and observe that the temperature dependence of $R$ in a low parallel magnetic field ($B_\parallel$) follows $R(T) \propto \exp\left(\left(\frac{T_0}{T}\right)^x\right)$, with $x = \frac{1}{2}$. This value is consistent with the variable-range hopping model of the Efros–Shklovskii type, indicating that the electrons are localized to form a Wigner crystal through strong Coulomb repulsive interaction[30]. As $B_\parallel$ increases to 12 T, the parameter $x$ becomes greater than 1/2 and finally reaches a value of around 0.7 (inset of Fig. 2e), which we attribute to the suppressed hopping of neighbouring localized electrons as their spins are polarized. Third, the resistance at $n = 7\frac{2}{3}n_0$ markedly increases when applying $B_\parallel$ (Fig. 2f) and is finally enlarged by approximately 20 times at $B_\parallel$ = 12 T (see Extended Data Fig. 1). This large increase in magnetoresistance again indicates that the effective inter-site hopping is hindered by a parallel magnetic field. Meanwhile, the magnetoresistance under a perpendicular magnetic field has a similar trend when $B_\perp$ < 7 T (Fig. 2g), which is a direct result of the weak spin–orbit coupling. Note that the giant positive magnetoresistance drops sharply at $B_\perp$ > 7 T (see details in Extended Data Fig. 2b), which is attributed to the dominance of the quantum Hall effect over the Wigner crystal state[31].

The interpretation of the Wigner crystal at the filling of $7\frac{2}{3}n_0$ in our cTDBG sample can be further supported by the small bandwidth and zero Chern number of the second moiré band according to theoretical calculations (see details in Methods). Thereby, we demonstrate prominent inter-site long-range Coulomb interactions in the triangular moiré lattice[23,27], showing that our cTDBG is an ideal platform to simulate the extended Hubbard model. In particular, the spin–valley isospin degrees of freedom existing in this platform would prompt exotic quantum many-body physics inaccessible in the simple SU(2) spin systems.

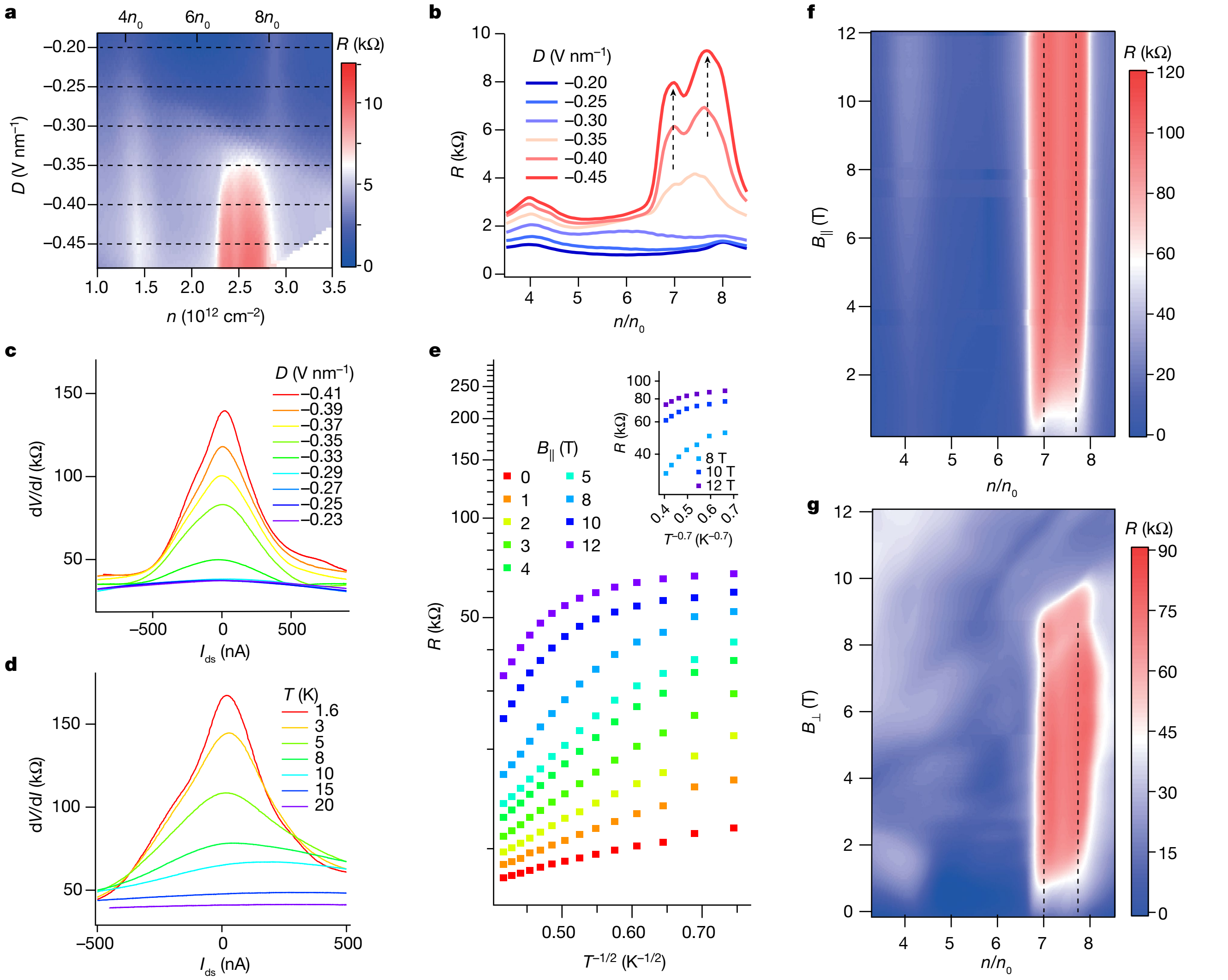

**Fig. 2 | Evidences for Wigner crystal state. a**, High-resolution 2D plot of $R$. Correlated insulating features emerge in the second moiré band. **b**, Line plots extracted from the dashed lines in **a** for selected displacement fields. The dashed arrows indicate peaks at $n = 7n_0$ and $7\frac{2}{3}n_0$. **c**,**d**, Differential resistance versus d.c. bias current at various displacement fields and temperatures, respectively. **e**, Resistance in log scale versus $T^{-\frac{1}{2}}$ at selected $B_{\parallel}$. Inset, resistance in log scale as a function of $T^{-0.7}$ at large $B_{\parallel}$. **f**,**g**, 2D map of $R$ as a function of the filling factor and parallel (perpendicular) magnetic field measured at $D = -0.46$ V nm$^{-1}$.

## Quantum two-stage criticality at zero magnetic field

The $D$-dependent electron correlations in the cTDBG enable the tuning of electron correlation and serve as a knob for investigating correlation-driven quantum criticality in the isospin extended Hubbard model. Figure 3a shows the temperature dependence of the resistance up to 20 K at a series of displacement fields, exhibiting three distinct regimes. At large $|D|$, as the temperature rises, the sheet resistance shows insulating behaviour until reaching a critical temperature at which the generalized Wigner crystal obtains enough thermal energy to melt into the metallic phase with linear $T$ dependence. At small $|D|$, the $T$-dependent resistance also shows linear behaviour at high temperatures but converts to $R \propto T^2$ when the temperature is lower than $T_n$ (see Methods for the determination of $T_n$). We note that, at low temperatures below the Fermi energy $E_F$, the resistivity in the 2D Fermi liquid phase has a temperature dependence $R \propto -(k_BT/E_F)^2 \ln\frac{k_BT}{E_F}$ (ref. [32]). Given that the logarithmic temperature-dependence term is hardly visible within a small temperature range such as that adopted in our measurements, the observed $R \propto T^2$ behaviour is a strong evidence for the normal Fermi liquid phase, in which the dominant scattering mechanism in the system is from electron–electron Coulomb interactions, rather than electron–phonon interactions. Moreover, at a series of intermediate $|D|$, the linear $T$ dependence of the resistance persists down to the base temperature.

To access the full phase diagram, we plot the 2D maps of $R$ and d$R$/d$T$ as functions of displacement field $D$ and temperature $T$, as shown in Fig. 3b,c. A fan-shaped regime, in which the resistance changes linearly with $T$, appears between the Wigner crystal phase and Fermi liquid phase, with the phase boundaries indicated by orange and red markers, respectively (see Methods for the algorithm to determine the boundaries). We refer to this $T$-linear regime as a strange metal phase, as it arises in the quantum critical regime[33,34]. Another smoking-gun signature of the strange metal is that the numerical pre-factor $C$ of the transport 'scattering rate' $\Gamma$ ($\Gamma = Ck_BT/\hbar$) should approach the Planckian dissipation limit[33,35], that is, $C = \frac{\hbar}{k_B}\frac{e^2 n_c}{m^*}\alpha \approx 1$. Here ħ is the Planck constant, $k_B$ is the Boltzmann constant, e is the unit charge, $n_c$ is the carrier density, $m^*$ is the effective mass and $\alpha$ is defined to fit $R = \alpha T + R_0$. In our $T$-linear regime, $n_c$ is the carrier density of the Wigner crystal with respect to the second moiré gap, that is, $n_c = \frac{1}{3}n_0 = 0.116 \times 10^{12}$ cm$^{-2}$, $m^* \approx 0.2m_e$ (see Methods) and $\alpha$ is approximately 92 Ω K$^{-1}$, yielding $C \approx 0.0126\alpha \approx 1.16$, which indeed approaches the Planckian dissipation limit.

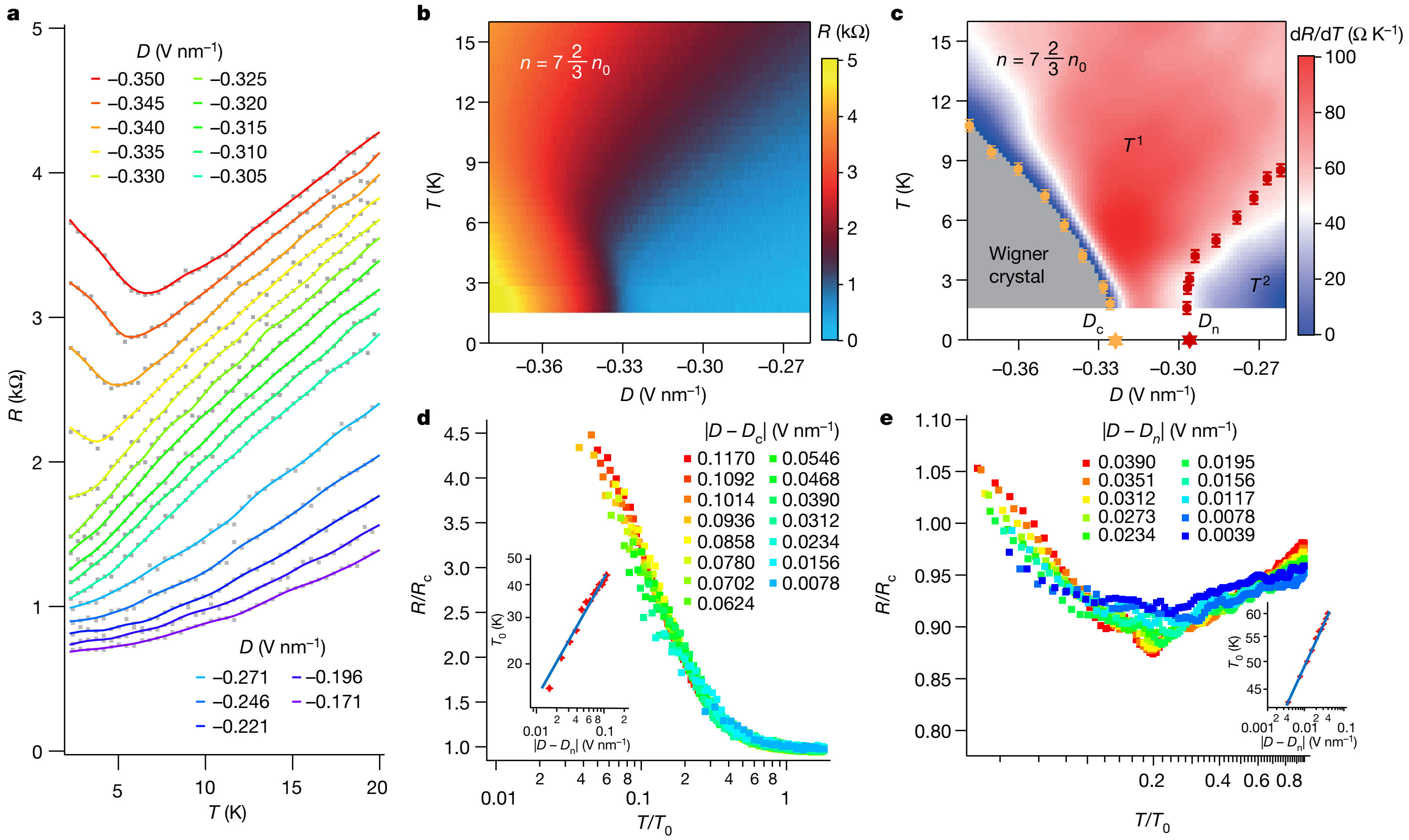

**Fig. 3 | Quantum two-stage criticality. a**, Line plots of the resistance for a series of $D$ ($R$–$T$ traces are offset for better clarification). **b**,**c**, 2D maps of $R$ and $dR/dT$ as functions of $D$ and $T$. Orange and red markers represent the phase boundaries, with error bars defined in Methods. $D_c$ and $D_n$ are the two quantum critical points. **d**,**e**, Scaling analysis for the quantum two-stage criticality in the Wigner crystal regime and the normal metal regime, respectively, with temperature scaling parameter $T_0$ chosen to yield collapse of the $R$–$T$ curves. Insets, $T_0$ versus $|D - D_c|$ and $|D - D_n|$, which follow power-law behaviours.

We perform quantum critical scaling analysis to gain further insight into the quantum critical behaviours. By normalizing the $D$-dependent $R$ ($T$) curves in the generalized Wigner crystal regime with the temperature-dependent resistance $R_c(T)$ at critical displacement field $D_c = -0.325$ V nm$^{-1}$, we observe that the normalized $R/R_c$ curves at different $D$ collapse onto a single branch, as shown in Fig. 3d. Moreover, the scaling parameter $T_0$ vanishes as the critical field is approached, following a power-law behaviour of $T_0 \sim |D - D_c|^{z\nu}$, with $z\nu = 0.50 \pm 0.02$ (inset of Fig. 3d). This power-law relation indicates that the QPT from the Wigner crystal to the strange metal is continuous. Notably, we observe a different quantum critical scaling behaviour in the normal metal regime by choosing critical field $D_n = -0.297$ V nm$^{-1}$, indicating another continuous QPT between the strange metal phase and the normal metal phase, as shown in Fig. 3e. The critical exponent is fitted to be $0.15 \pm 0.01$ (inset of Fig. 3e) and is much smaller than that extracted from the scaling analysis at $D_c$, indicating that these two QPTs belong to different universality classes. Moreover, we show that valid scaling analysis cannot be performed at any displacement field between −0.297 V nm$^{-1}$ and −0.325 V nm$^{-1}$ (Extended Data Fig. 3). This means that the critical points of these two QPTs are separated by a window of $D$, which suggests the existence of a critical intermediate phase between the Wigner crystal and the Fermi liquid. We term this phenomenon of two quantum critical points with a critical intermediate phase as a quantum two-stage criticality.

## Quantum pseudo criticality at high parallel magnetic field

The decoupled spin and valley degrees of freedom in our isospin extended Hubbard simulator enables the capability to polarize the spins without affecting the valley degree of freedom by applying a parallel magnetic field $B_\parallel$, thus giving a unique opportunity to shed light on the interplays between the internal degrees of freedom and spatial charge orders. We first set $B_\parallel$=1.5 T and 3 T and extract the $D$–$T$ phase diagrams, as shown in Extended Data Fig. 4. Combining with the result at zero $B_\parallel$ (Fig. 3c), we observe that, when $B_\parallel$ increases, the critical displacement field $D_c$ increases, showing an enlarged regime of the generalized Wigner crystal. By contrast, the fan-shaped strange metal regime shrinks and the $D$ window between the two critical points, that is, $D_c$ and $D_n$, gradually vanishes as $B_\parallel$ increases.

We further increase $B_\parallel$ to 12 T and measure the temperature-dependent resistance at a series of displacement fields, as shown in Fig. 4a. We then plot the map of $dR/dT$ as a function of $D$ and $T$ in Fig. 4b, showing that the quantum critical regime of the strange metal phase only exists above a critical temperature $T^* \approx 5.6$ K, as indicated by the yellow star. We note that this temperature is equivalent to the spin Zeeman energy in a magnetic field of about 8.3 T, which is comparable with the $B_\parallel$ (12 T) in our measurements, indicating that the spins are frozen below $T^*$. Notably, at temperatures below $T^*$, we identify a range of $D$ in which $R$ is almost independent of $T$, as shown in the dashed box in Fig. 4a.

The unconventional phase diagram in Fig. 4b prompts us to investigate the quantum criticality therein by performing the scaling analysis. However, we find that no quantum critical point can be identified for a reasonable $R$–$T$ curve collapse if we carry out the scaling for the full temperature range, that is, from 1.5 K to 30 K. Notably, for $R$–$T$ curves in the insulating regime above $T^*$, they all collapse onto a single branch after performing scaling analysis with the critical point $D_c^* = -0.273$ V nm$^{-1}$, as shown in Fig. 4c. Similarly, successful and failed collapse of scaled $R$–$T$ curves with the critical point $D_n^* = -0.283$ V nm$^{-1}$ are observed in the normal metal regime for temperatures above $T^*$ only and for the full temperature range, respectively (Extended Data Fig. 5).

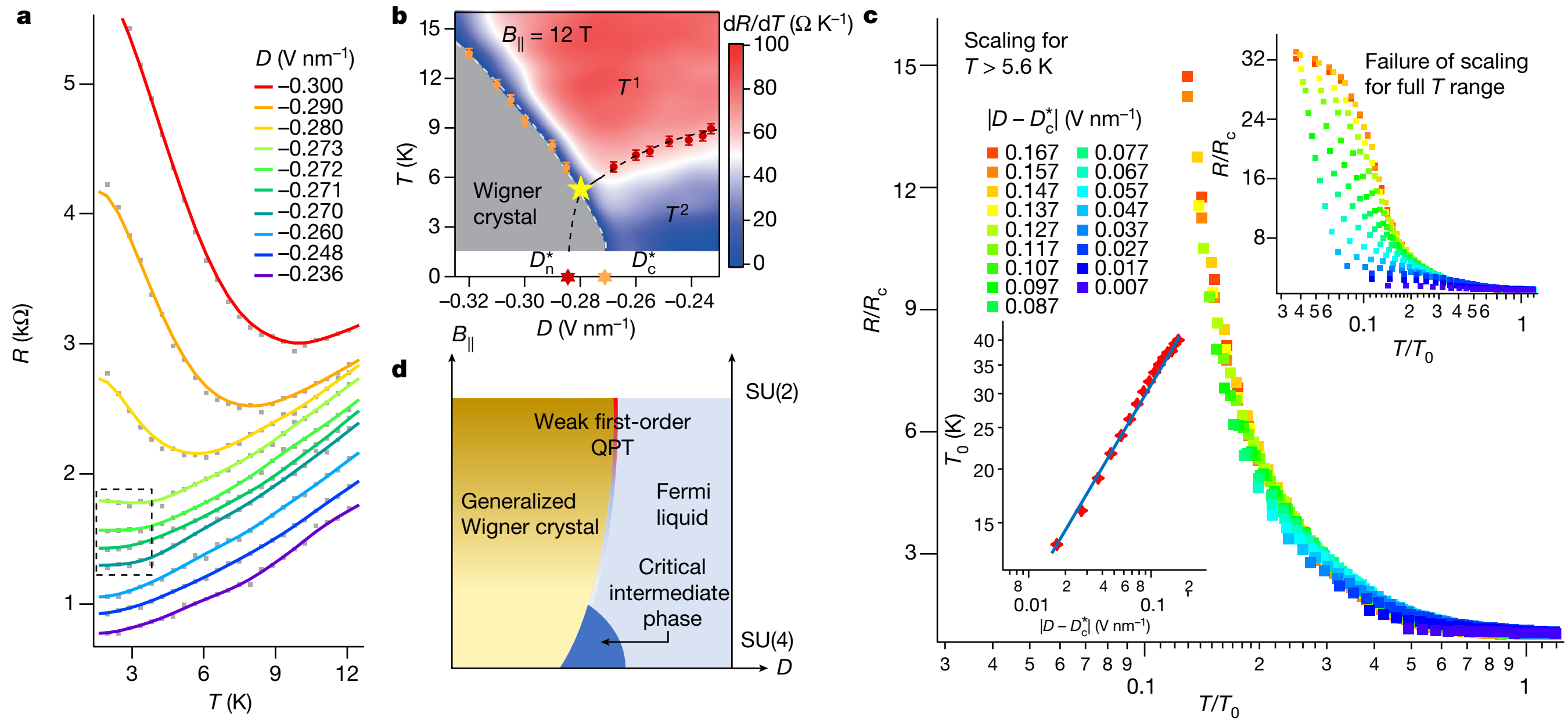

**Fig. 4 | Quantum pseudo criticality in a 12-T parallel magnetic field. a**, Line plots of $R$–$T$ curves at different $D$ (the curves are offset for better clarification). The dashed box indicates the almost-flat $R$–$T$ curves. **b**, 2D map of d$R$/d$T$ as a function of $D$ and $T$ for $B_{\parallel} = 12$ T. **c**, Successful (lower inset) and failed (upper inset) collapse of $R$–$T$ curves by performing the scaling analysis in the Wigner crystal regime for temperatures above $T^*$ only and for the full temperature range, indicating the emergence of a quantum pseudo criticality. **d**, Schematic of phase diagram and quantum phase transitions in the plane of $D - B_{\parallel}$.

These observations indicate the emergence of a quantum pseudo criticality[36,37]. By contrast to $D_c < D_n$ at zero $B_{\parallel}$, we observe $D_c^* > D_n^*$ at $B_{\parallel} = 12$ T, indicating an overlapped regime instead of a critical intermediate phase. We also note that the critical scaling exponent $\upsilon z \approx 0.5$ in the Wigner crystal regime is the same as that at $B_{\parallel} = 0$, whereas the extracted $\upsilon z$ in the normal metal regime increases from 0.15 at $B_{\parallel} = 0$ to 0.34 at $B_{\parallel} = 12$ T.

## Discussion

Although the strange metal behaviour near quantum criticality previously observed in graphene moiré systems has been attributed to the quantum fluctuations of the valley[35], in our work, the strange metal phase shrinks as the spins are polarized by the parallel magnetic field, indicating that the quantum fluctuations of spins play the dominant role. Besides, the intermediate ground state is spin unpolarized because the increased $B_{\parallel}$ gradually narrows and eventually closes the $D$ window that accommodates the intermediate ground state. To get insight into the dominant physics of this intermediate ground state, we carry out the mean-field-theory analysis and identify charge density wave with valence bond state (VBS) as a candidate (see details in Section VI of the Supplementary Materials). At finite $T$, the fluctuations of the VBS may couple to the charge degrees of freedom and lead to the breakdown of well-defined coherent quasiparticles, resulting in the $T$-linear resistivity in our observations. Note that the spatial and quantum fluctuations are not involved in our mean-field calculations. Other numerical methods beyond mean-field theory are required to provide comprehensive understanding of the intermediate phase and its $B_{\parallel}$ dependence in the future.

Our observation of quantum pseudo criticality at high parallel magnetic field indicates that the transition between spin-polarized generalized Wigner crystal and Fermi liquid is a weak first-order QPT. The almost $T$-independent $R$ and failure of scaling analysis below $T^*$ suggest the existence of a small energy scale $\Delta^* = k_B T^*$, which equals the spin Zeeman energy in our case, and—equivalently—a large but finite correlation length $\xi$ (refs. [36,37]). For temperatures $T > T^*$, the thermal energy will 'melt' the frozen spins in high parallel magnetic field, and the system still behaves as if critical. Notably, the weak first-order QPT has been intensively discussed in the context of deconfined quantum critical point[38,39], in which Néel–VBS transition for SU(2) spins is argued to be a weak first-order QPT instead of a continuous one. Moreover, a recent theory has suggested a close connection of deconfined quantum critical point and the metal–insulator transition between generalized Wigner crystals and Fermi liquid in the SU(2) spin system[40]. Our findings help build up a solid-state platform simulating an evolution from QPT with critical intermediate phase in the high-symmetric SU(4) system to a weak first-order QPT in the low-symmetric SU(2) system (Fig. 4d), and would offer a new perspective to understand the strongly correlated physics in the electron systems with realistic SU(2) spins.

## Online content

Any methods, additional references, Nature Research reporting summaries, source data, extended data, supplementary information, acknowledgements, peer review information; details of author contributions and competing interests; and statements of data and code availability are available at https://doi.org/10.1038/s41586-022-05106-0.

## Methods

### Device fabrication

We fabricated the twisted double bilayer graphene by a modified polymer-based cut-pick technique. We mechanically exfoliated bilayer graphene, graphite and hexagonal boron nitride (h-BN) onto a highly doped Si wafer covered by a 300-nm-thick $SiO_2$ layer, with the thickness of these flakes identified with optical contrast and a Bruker MultiMode 8 atomic force microscope. The bilayer graphene was cut into two halves by using the atomic force microscope tip in contact mode. We then used stamps consisting of poly(bisphenol A carbonate) film and polydimethylsiloxane to pick up top graphite and h-BN at around 80 °C. After that, we picked up one half of the bilayer graphene and then the other half with a rotation of 180°+0.8° at room temperature. This aimed rotation angle is slightly larger than the twisted angle of our measured device, which is attributed to the inevitable angle relaxation in the fabrication process. Finally, we picked up bottom h-BN at 80 °C and the five-layer heterostructures were transferred onto bottom graphite at around 135 °C. The heterostructures were then shaped into Hall bar geometry by using standard electron-beam lithography and dry etching in an inductively coupled plasma system. The edge contact electrodes (10 nm Cr/10 nm Pd/30 nm Au) to cTDBG and top electrodes (5 nm Ti/40 nm Au) to the top and bottom graphite gates were deposited by standard electron-beam evaporation. The temperature was always kept below 150 °C during stacking and fabrication processes.

### Electrical measurements

The devices were measured in an Oxford cryostat with a base temperature of about 1.5 K. The resistance signals were collected using a low-frequency (17.7 Hz) SR830 Lock-in Amplifier (Stanford Research). Gate voltages were applied with Keithley 2400 Source Meters. The differential resistance signals were collected by applying an a.c. excitation current of 1 nA added on top of a variable d.c. bias current up to 1 μA. We apply d.c. currents and record amplified voltages by adopting a data acquisition system (National Instruments 6251).

### Continuum model of twisted double bilayer graphene

We consider AB-BA-stacked twisted double bilayer graphene: an AB-stacked bilayer graphene is placed on top of another BA-stacked bilayer graphene and they are twisted with respect to each other by an angle $\theta$, forming a moiré superlattice. The moiré lattice constant $L_s = a/2\sin(\theta/2)$, in which $a = 0.246$ nm, is the atomic lattice constant of monolayer graphene. We further consider atomic corrugation in the system, that is, the interlayer distance $d(\mathbf{r})$ between the two twisted layers at the interface (between the AB and BA bilayers) varies as a function of in-plane position $\mathbf{r}$ in the moiré supercell, which can be approximately modelled as $d(\mathbf{r}) = d_0(\mathbf{r}) + 2d_1\sum_{i=1}^{3}\cos\mathbf{g}_i\cdot\mathbf{r}$, in which $\mathbf{g}_1$, $\mathbf{g}_2$ and $\mathbf{g}_3 = \mathbf{g}_1 + \mathbf{g}_2$ are the three reciprocal lattice vectors of the moiré supercell. We take $d_0 = 0.3433$ nm and $d_1 = 0.00278$ nm to reproduce the interlayer distances of the AA-stacked and AB-stacked bilayer graphene at the ABBA point and the ABAC point in the moiré supercell, respectively.

The low-energy electronic structure of cTDBG can be well described by the Bistritzer–MacDonald continuum model[41], in which the low-energy states from the two atomic valleys $K$ and $K'$ are assumed to be completely decoupled from each other. To be specific, the continuum model for the $K$ valley of the cTDBG system is expressed as

$$\mathrm{H}^{K}_{\mathrm{AB-BA}} = \begin{pmatrix} \mathrm{H}^{K}_{\mathrm{AB}} & \mathbb{U} \\ \mathbb{U}^{\dagger} & \mathrm{H}^{K}_{\mathrm{BA}} \end{pmatrix} \tag{1}$$

in which $\mathrm{H}^{K}_{\mathrm{AB}}$ and $\mathrm{H}^{K}_{\mathrm{BA}}$ are the Hamiltonians of the untwisted bilayers:

$$\mathrm{H}^{K}_{\mathrm{AB}} = \begin{pmatrix} -\hbar v_{\mathrm{F}}(\mathbf{k}-\mathbf{K_1})\cdot\sigma & h_+ \\ h_- & -\hbar v_{\mathrm{F}}(\mathbf{k}-\mathbf{K_1})\cdot\sigma \end{pmatrix} \tag{2}$$

and

$$\mathrm{H}^{K}_{\mathrm{BA}} = \begin{pmatrix} -\hbar v_{\mathrm{F}}(\mathbf{k}-\mathbf{K_2})\cdot\sigma & h_- \\ h_+ & -\hbar v_{\mathrm{F}}(\mathbf{k}-\mathbf{K_2})\cdot\sigma \end{pmatrix} \tag{3}$$

in which $\mathbf{K_1}$ ($\mathbf{K_2}$) denotes the Dirac point of the AB-stacked (BA-stacked) bilayer. The Pauli matrices $\sigma = (-\sigma_x, \sigma_y)$ are defined in the space of the A, B sublattices of graphene. $h_+$ denotes the interlayer hopping matrix between the untwisted bilayer,

$$h_+ = \begin{pmatrix} t_2 f(\mathbf{k}) & t_2 f^*(\mathbf{k}) \\ t_\perp - 3t_3 & t_2 f(\mathbf{k}) \end{pmatrix} \tag{4}$$

in which $t_\perp = 0.48$ eV, $t_2 = 0.21$ eV and $t_3 = 0.05$ eV are the nearest-neighbour, second-neighbour and third-neighbour interlayer hopping amplitudes, respectively, and $h_- = h_+^{\dagger}$.

We consider only the nearest-neighbour interlayer coupling between the two twisted bilayers, that is, the coupling between the top layer of the AB bilayer and the bottom layer of the BA bilayer, which is expressed as

$$\mathbb{U} = \begin{pmatrix} 0 & 0 \\ U(\mathbf{r})\mathrm{e}^{-\mathrm{i}\Delta\mathbf{K}\cdot\mathbf{r}} & 0 \end{pmatrix} \tag{5}$$

in which $\Delta\mathbf{K} = \mathbf{K}_2 - \mathbf{K}_1 = (0, 4\pi/3L_s)$ is the shift between the two Dirac points of the two sets of twisted bilayers[42]. $U$ is the interlayer hopping term between the two Dirac states of the twisted layers[43],

$$U = \begin{pmatrix} u_0 g(\mathbf{r}) & u_0' g(\mathbf{r}-\mathbf{r}_{\mathrm{AB}}) \\ u_0' g(\mathbf{r}-\mathbf{r}_{\mathrm{AB}}) & u_0 g(\mathbf{r}) \end{pmatrix} \tag{6}$$

in which $\mathbf{r}_{\mathrm{AB}} = (\sqrt{3}L_s/3,\ 0)$, $g(\mathbf{r}) = \sum_{j=1}^{3}\mathrm{e}^{\mathrm{i}\mathbf{q}_j\cdot\mathbf{r}}$, with $\mathbf{q}_1 = (0, 4\pi/3L_s)$, $\mathbf{q}_2 = (-2\pi/\sqrt{3}L_s, -2\pi/3L_s)$ and $\mathbf{q}_3 = (2\pi/\sqrt{3}L_s, -2\pi/3L_s)$ (ref. [44]). In cTDBG, the corrugation effects result in $u_0 < u_0'$, in which $u_0 \approx 0.078$ eV and $u_0' \approx 0.097$ eV are the intra-sublattice and inter-sublattice interlayer coupling terms, respectively[43]. We note that all the parameters of the continuum model, as given in equations (1)–(6), are derived from a realistic Slater–Koster tight-binding model introduced in ref. [45]. We have also compared the results calculated from the continuum model with those calculated using the atomic Slater–Koster tight-binding model and find excellent agreement with each other.

The displacement field $D$ is introduced by applying a homogeneous vertical electrostatic potential drop across the four layers

$$\mathrm{H}^{K}_{\mathrm{AB-BA}} = \begin{pmatrix} h_0(\mathbf{k}) - U_{\mathrm{d}}/2 & h_+ & 0 & 0 \\ h_- & h_0(\mathbf{k}) - U_{\mathrm{d}}/6 & U(\mathbf{r})\mathrm{e}^{-\Delta\mathbf{K}\cdot\mathbf{r}} & 0 \\ 0 & U^{\dagger}(\mathbf{r})\mathrm{e}^{\Delta\mathbf{K}\cdot\mathbf{r}} & h_0(\mathbf{k}) + U_{\mathrm{d}}/6 & h_- \\ 0 & 0 & h_+ & h_0(\mathbf{k}) + U_{\mathrm{d}}/2 \end{pmatrix} \tag{7}$$

in which $U_{\mathrm{d}} = \mathrm{e}D\cdot d_t/\epsilon$ is the electrostatic potential energy difference between the top and bottom layers, $D$ is the displacement field, $d_t \approx 1.005$ nm is the thickness of the entire system and $\epsilon \approx 4$ is the dielectric constant of the BN substrate. We note that several different values for the dielectric constant of h-BN has been widely used, which can lead to markedly different conclusions in the theoretical calculations. In our experiment, the adoption of $\epsilon \approx 3.8$ gives us a consistent result, so we choose the value of 4 for our theoretical calculations.

We note that the Coulomb potentials from the fully occupied bands below the target flat band may strongly enhance the bandwidth of the flat band and renormalize the flat-band wavefunction. This question has recently attracted a lot of attention. For example, Vafek and Kang recently performed comprehensive renormalization group studies on magic-angle twisted bilayer graphene[46] and found that both the kinetic energy and the Coulomb interaction energy increase as the Fermi level approaches the charge neutrality point. In the end, when the flat bands are partially occupied, the ratio between the interaction energy projected to the (renormalized) flat-band subspace and the renormalized bandwidth is roughly unchanged, and the system is still in the strong-coupling regime. A similar argument could also be applied to the twisted double bilayer graphene system.

To calculate the numerical prefactor $C$ of the transport 'scattering rate', we have used the $m_e$ at the valence band edge (in which the effective mass can be well defined) and the carrier density corresponding to the $8n_0$ filling. Because the 7+2/3 filling in our device is in very close proximity to the $8n_0$ filling, it is therefore reasonable that the $C$ value calculated for the $8n_0$ filling applies to that at the 7+2/3 filling.

### Criterion for the Wigner crystal state

Here we introduce a dimensionless quantity $r_s = V/W$, the ratio between the Coulomb interaction energy ($V$) and kinetic energy ($W$) of the electrons, as the key criteria for the formation of Wigner crystal state[47]. The average distance between the electrons in the moiré bands is denoted as $r_e$, which satisfies $\pi r_e^2 = 1/n_e$, in which $n_e = \upsilon/\Omega$ is the carrier density, with $\Omega$ the area of the moiré primitive cell and $\nu$ the filling factor. Then the characteristic Coulomb interaction energy is $V = e^2/4\pi\epsilon_0\epsilon r_e$, in which e is the elementary electronic charge, $\epsilon \approx 4$ is the dimensionless background dielectric constant and $\epsilon_0$ is the vacuum permittivity. The characteristic kinetic energy of the system can be effectively described by the free electron kinetic energy $W = \hbar^2 k_F^2/2m_e^*$, in which $m_e^*$ is the effective mass of the carrier at the Fermi level and $k_F$ is the Fermi wave vector. Notably, $m_e^*$ is calculated to be about $0.2m_e$, based on the formula $\frac{1}{m^*} = \frac{1}{\hbar^2}\nabla_{\mathbf{k}}^2 E(\mathbf{k})$. The carrier density is related to the Fermi wave vector:

$$n_e = 4\int \frac{d^2p}{(2\pi)^2} n_k = 4\int_0^{k_F} \frac{p\cdot dp}{2\pi} = \frac{k_F^2}{\pi} \tag{8}$$

in which the factor of 4 is from the fourfold valley–spin degeneracy. Then we obtain:

$$r_s = \frac{2}{\epsilon}\frac{m_e^*}{m_0}\frac{r_e}{a_B} \tag{9}$$

in which $a_B = 4\pi\epsilon_0\hbar^2/m_0e^2$ is the Bohr radius and $m_0$ is the bare electron mass. According to previous studies[48,49], electrons crystallize for $r_s \geq 31$ in the one-valley two-dimensional electron gas (2DEG) system and for $r_s \geq 30$ for the two-valley 2DEG system[49].

We note that the small bandwidth of the second conduction flat band (around 6–8 meV) gives rise to $r_s$ of about 35 (see Extended Data Fig. 6a), which exceeds the value for Wigner crystallization in 2DEGs[49]. The zero Chern number indicates a trivial band topology different from the topological charge density wave and fractional Chern insulator residing in the first moiré band of other moiré graphene systems[50–52]. Notably, the Chern number is non-zero in twisted double bilayer graphene of the stacking of AB-AB at the same fillings and $D$ field, indicating that the stacking order plays a crucial role in forming the Wigner crystal. In addition, the Wigner crystal state only emerges at a large displacement field, at which there is only one Fermi surface with large $r_s$ centred at the $\Gamma_s$ point for each valley (see Extended Data Fig. 6b). By contrast, as $|D|$ is reduced to zero, other Fermi surfaces with much smaller $r_s$ emerge (see Extended Data Fig. 6c,d), making the system still behave as a Fermi liquid.

### Scaling analysis

The procedure of scaling analysis is detailed described next. (1) We first measure the resistance $R$ at different temperatures $T$ and displacement fields $D$, obtaining several $R$–$T$ curves at different $D$, with typical results shown in Fig. 3a. (2) We choose a trial critical field $D^*$ near the phase boundaries at the base temperature in the measurements, which has $R$–$T$ curve denoted by $R^*(T)$, and normalize the $R$–$T$ curves at different $D$ in the Wigner crystal regime by $R^*(T)$. We then try to make the normalized $R/R^*(T)$ curves at different $D$ collapse onto a single branch after scaling the temperatures by $D$-dependent $T_0$. If this fails, then find a nearby point to repeat this process and find a point that has the optimal collapse of $R$–$T$ curves. The critical $D_c$ for the Wigner crystal regime is determined to be −0.325 V nm$^{-1}$ in our device in zero magnetic field. A similar method has been used for other scaling analyses in this work.

### Determination of $T_n$

Now we explain in detail the algorithm for how we determine the phase boundary $T_n$ between the strange metal and normal metal phases. In the normal metal region ($T < T_n$), the $R$–$T$ relation follows $R(T) = AT^2 + R_0$, whereas in the strange metal region ($T > T_n$), $R$ is proportional to $T$. It follows that the temperature boundary $T_n$ between strange metal and normal metal can be determined by fitting the functional relationship of the resistance $R$ and temperature $T$. We first attempt to fit $R$ to the form $R(T) = AT^2 + R_0$ from the base temperature to a trial maximum temperature ($T_{trial}$) and extract the $r_p^2$ representing the correlation coefficient of such parabola fitting. Meanwhile, we fit $R$ to the form $R(T) = AT + R_0$ for $T > T_{trial}$ to extract another $r_l^2$ representing the correlation coefficient of this linear fitting. Then we try several different $T_{trial}$ for the fitting to search for the temperature simultaneously having high $r_p^2$ and $r_l^2$ values, which is then determined as $T_n$. Extended Data Fig. 7 shows a typical example of fitting at $n = 7\frac{2}{3}n_0$ and $D = -0.21$ V nm$^{-1}$. We find that the resistance is fit well ($r^2 = 0.98$) by a $T^2$ form up to a temperature of 14.7 K and is also fit well ($r^2 = 0.996$) by a $T$-linear form between 14.7 and 30 K. Here we can also determine the error of the fitting. If there exist a few trial temperatures that all give good fitting results, then we can define this range of $T$ as the error of fitting. If there is only one trial temperature that can give good fitting, then we define the error by the step length of the temperature in our measurements. Such error analysis gave the error bars in Figs. 3c and 4b.

## Data availability

The data that support the plots in this paper and other findings of this study are available from the corresponding authors on reasonable request.

**Acknowledgements** This work was supported in part by the National Natural Science Foundation of China (62034004, 62122036, 61921005, 12074176, 61974176 and 11874205) and the Strategic Priority Research Program of the Chinese Academy of Sciences (XDB44000000). J.L. was supported by National Key R&D program of China (grant no. 2020YFA0309601) and the start-up grant of ShanghaiTech University. C.W. was supported by the Research Grant Council of Hong Kong (CRF C7012-21GF). K.W. and T.T. acknowledge support from JSPS KAKENHI (grant numbers 19H05790, 20H00354 and 21H05233) and A3 Foresight by JSPS. Q.L. would like to thank M. Wang for valuable discussions.

**Author contributions** F.M., B.C. and Q.L. conceived the idea and designed the experiments. F.M. and B.C. supervised the whole project. Q.L. fabricated cTDBG devices and performed the measurements. Q.L. and B.C. analysed the experimental data. M.C., Y.X., P.W., F.C. and Z.L. provided assistance in the experiments. B.X., J.L. and D.W. developed the theoretical model. T.T. and K.W. provided h-BN samples. Q.L., B.C. and F.M. co-wrote the manuscript, with extensive input from J.L., Q.-H.W., S.-J.L., D.W. and C.W.

**Competing interests** The authors declare no competing interests.

**Additional information**

**Correspondence and requests for materials** should be addressed to Bin Cheng or Feng Miao.

**Peer review information** *Nature* thanks Mandar Deshmukh and the other, anonymous, reviewer(s) for their contribution to the peer review of this work.

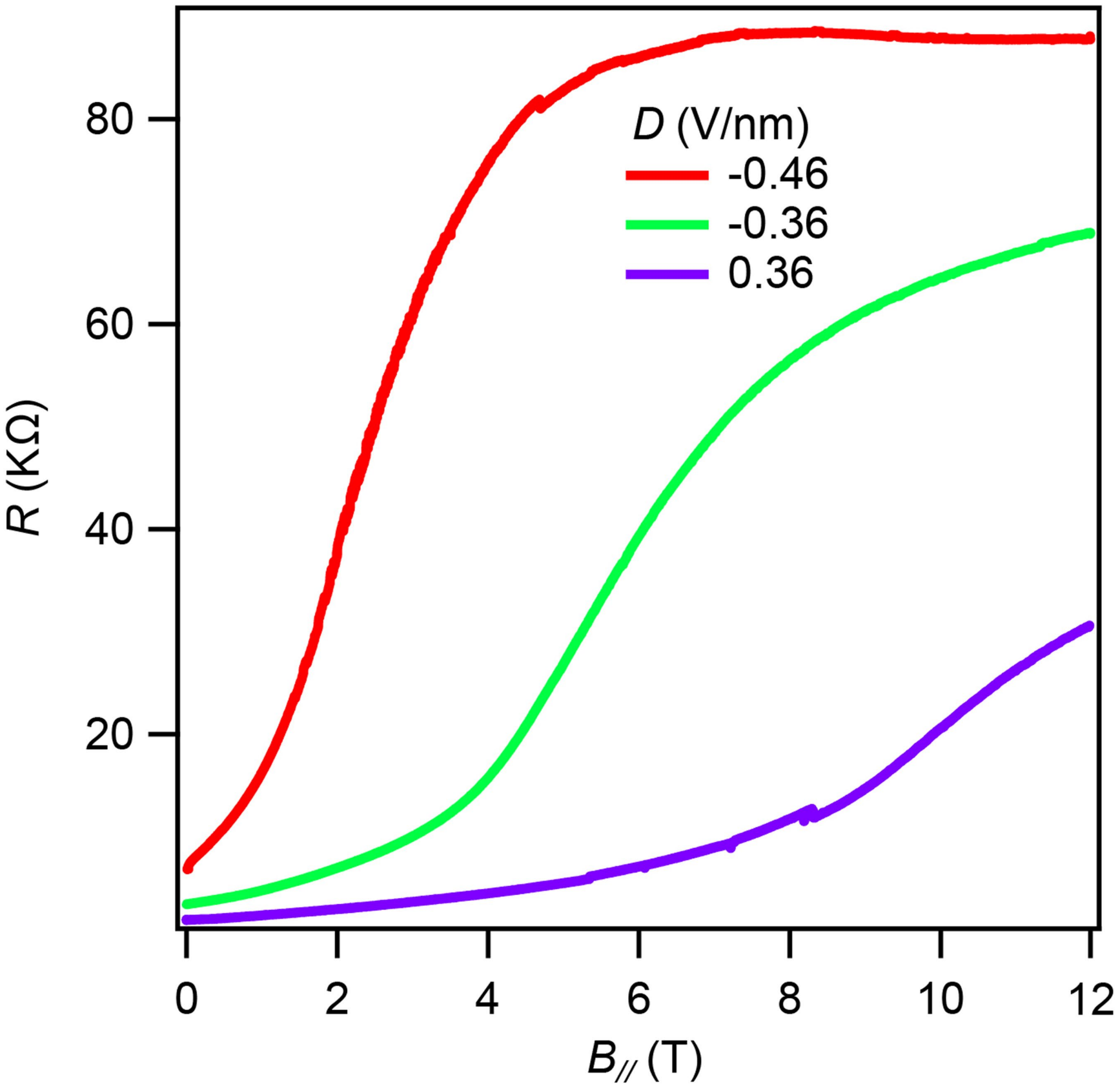

**Extended Data Fig. 1 | Parallel magnetic field dependence of resistance at different *D*.** Resistance as a function of parallel magnetic field at selected displacement fields. The saturation of resistance at high parallel magnetic field suggests that the spins of the localized electrons are totally polarized.

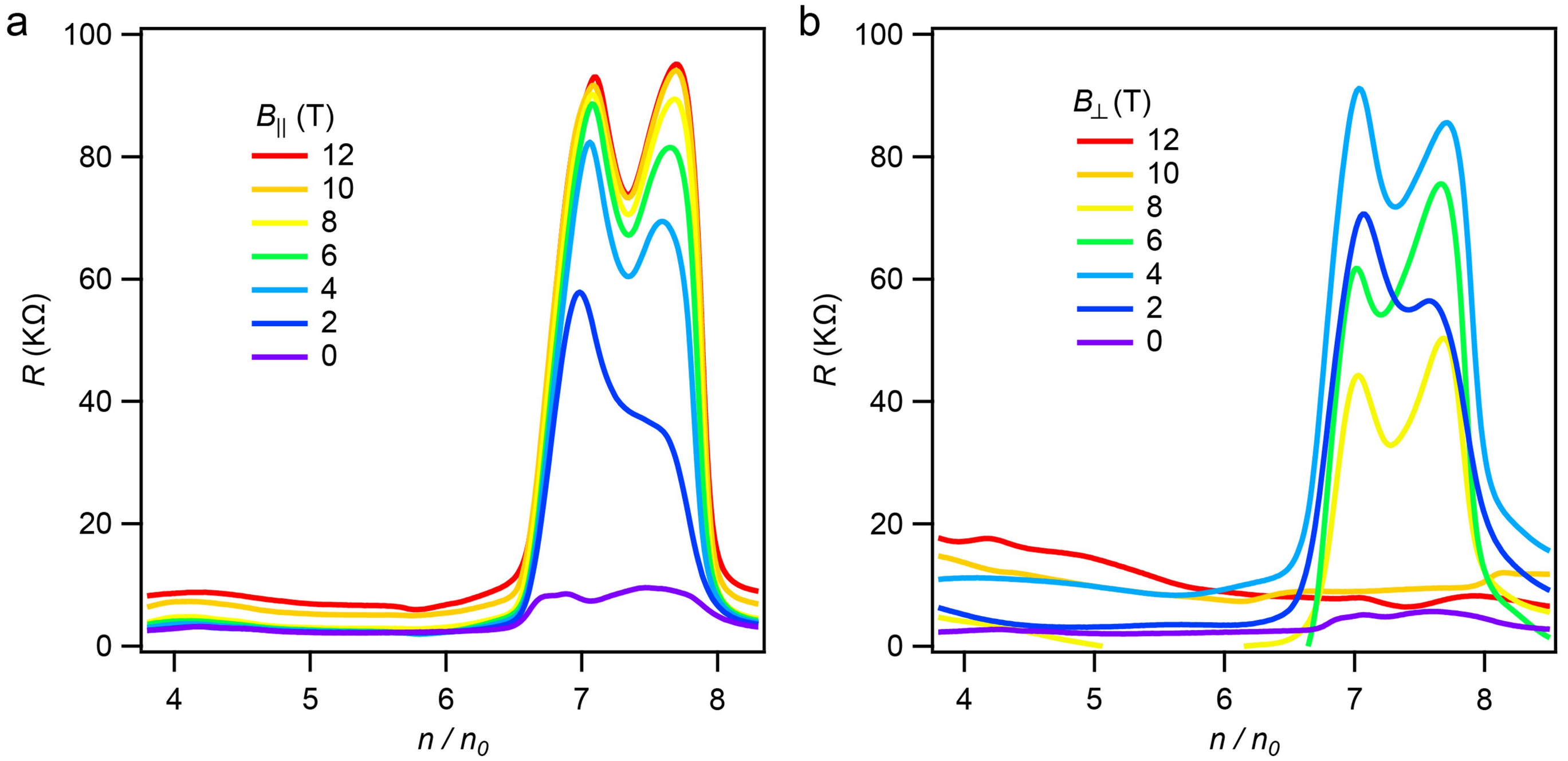

**Extended Data Fig. 2 | Line traces extracted from Fig. 2f,g at several magnetic fields. a**, As parallel magnetic field increases, resistance at fractional filling $n = 7\frac{2}{3}n_0$ first increases monotonically and gradually saturates. **b**, For the case of vertical magnetic field, resistance at fractional filling $n = 7\frac{2}{3}n_0$ increases first and is finally overwhelmed by quantum Hall effect at high $B_{\perp}$.

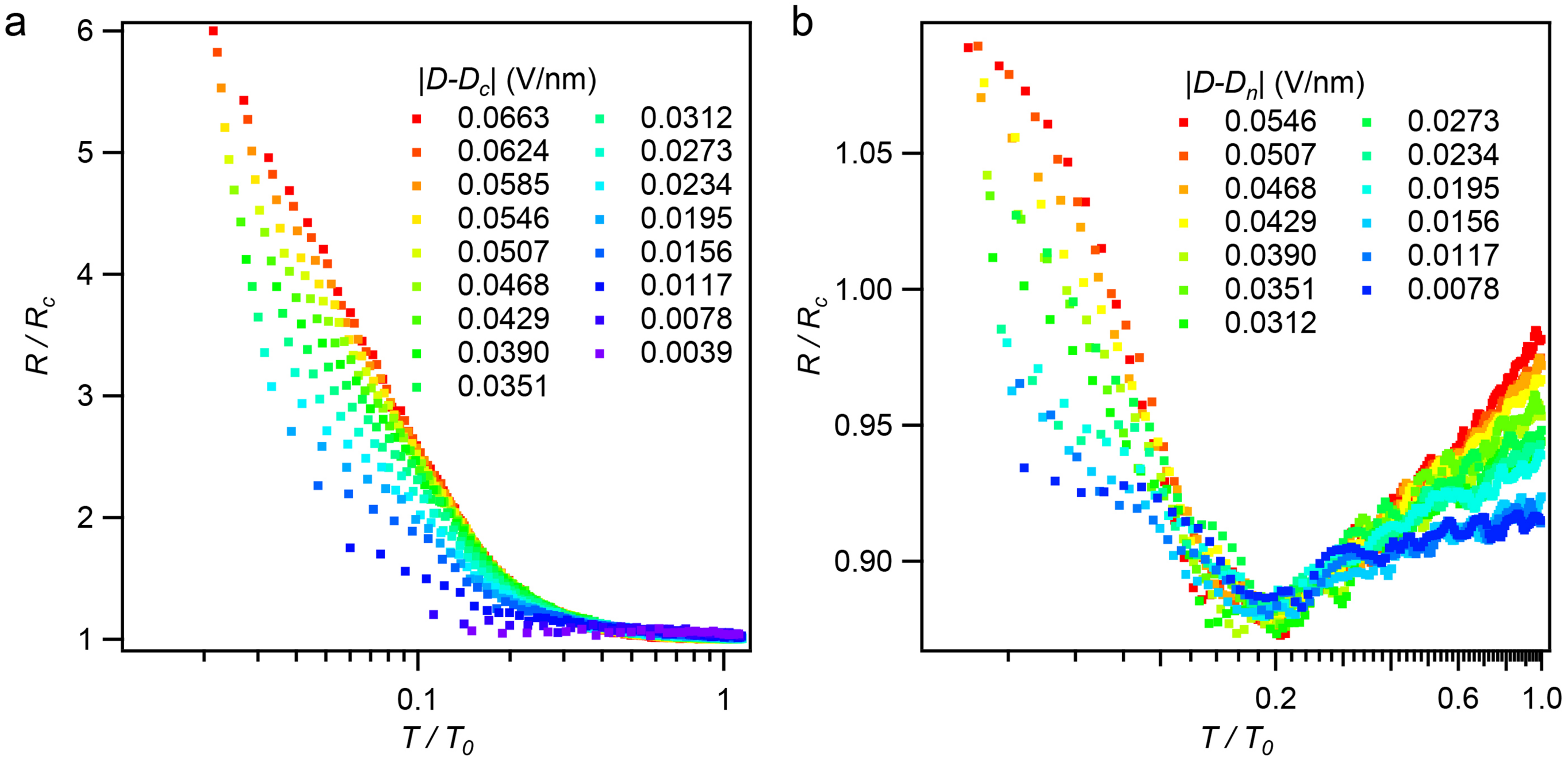

**Extended Data Fig. 3 | Failure of scaling analysis for Wigner crystal and normal metal by choosing $D_c$ and $D_n$ within the intermediate regime.** Here we show two typical failures of scaling. **a**, Scaling analysis for Wigner crystal by choosing $D_c = -0.316$ V nm$^{-1}$. The curves are unable to collapse onto a single curve compared with the scaling analysis in the main text with $D_c = -0.325$ V nm$^{-1}$. **b**, Scaling analysis is unable to collapse the $R$–$T$ curves if we choose $D_n = -0.305$ V nm$^{-1}$ as the critical displacement field for the normal metal regime.

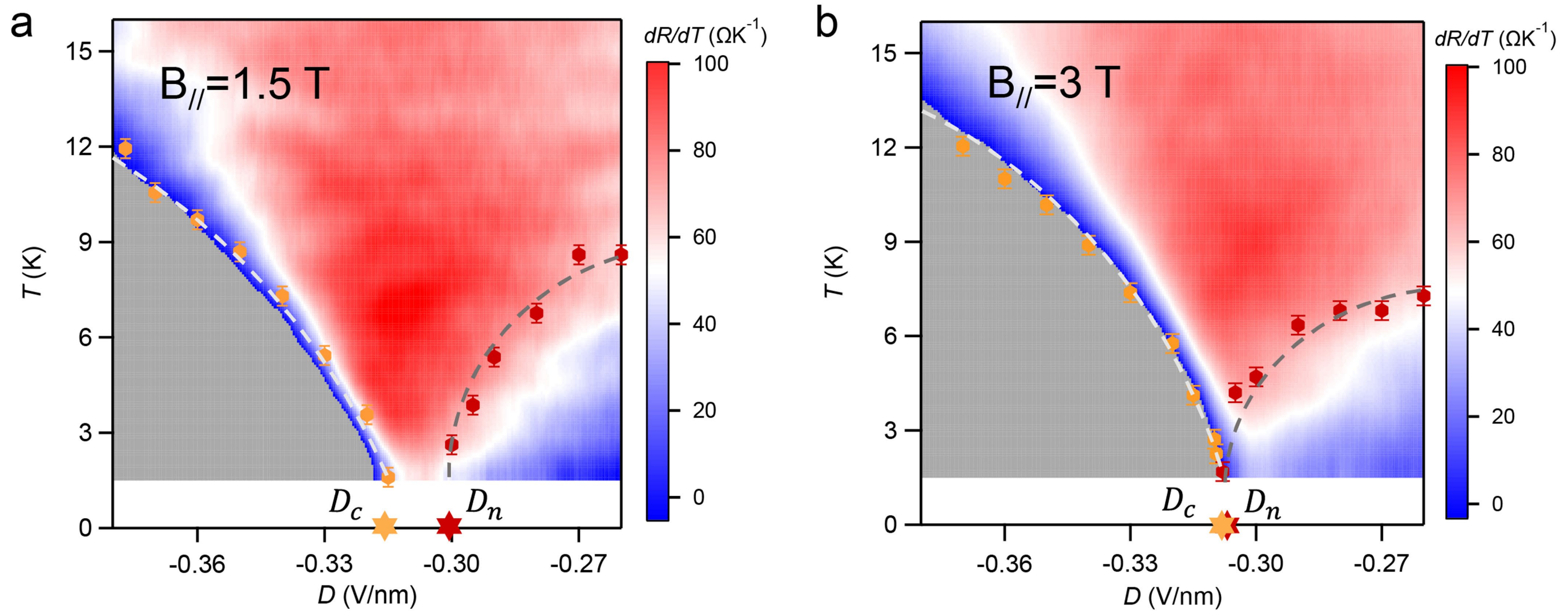

**Extended Data Fig. 4 | Evolution of the quantum critical behaviour with parallel magnetic field.** 2D map of d$R$/d$T$ as a function of $D$ and $T$ at $B_{||}$ = 1.5 T (**a**) and 3 T (**b**).

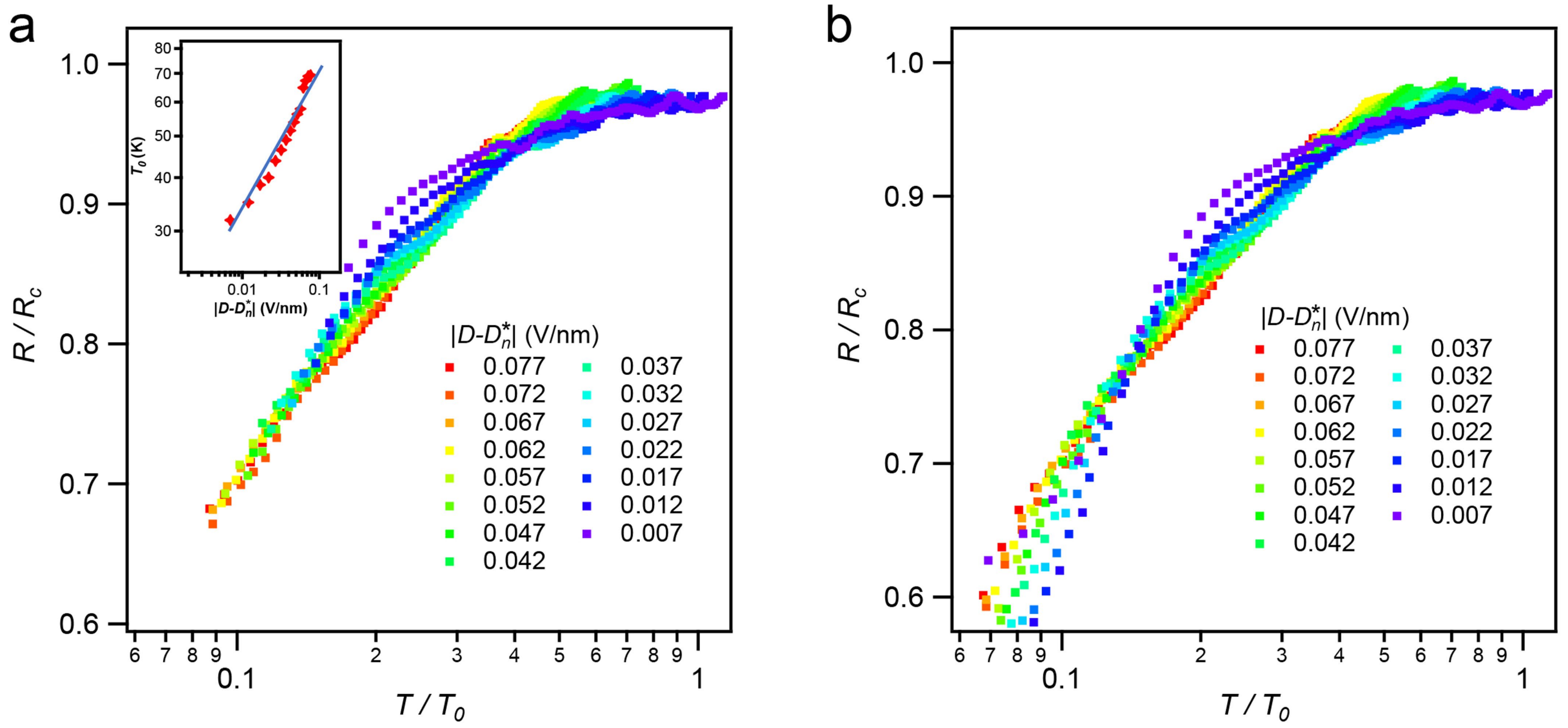

**Extended Data Fig. 5 | Quantum pseudo criticality in the normal metal regime.** Successful (**a**) and failed (**b**) collapse of scaled *R*–*T* curves in the normal metal regime for temperatures above $T^*$ only and for the full temperature range.

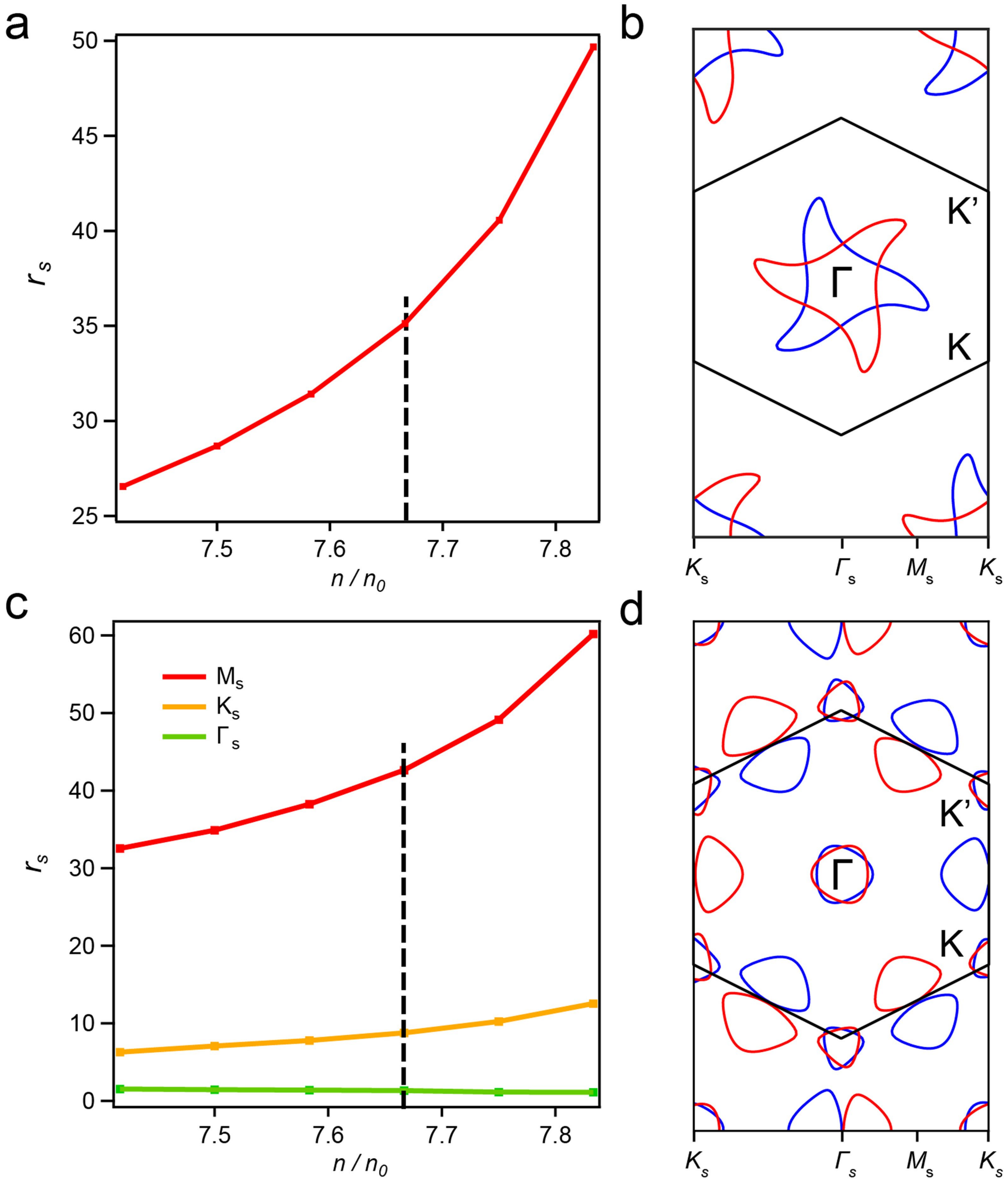

**Extended Data Fig. 6 | $r_s$ evolution with varying filling factor and Fermi surfaces at $n = 7\frac{2}{3}n_0$ at different $D$.** $r_s$ of AB-BA-stacked graphene under $D = -0.4$ V nm$^{-1}$ from hole pockets near $\Gamma_s$ points (**a**) and $D = 0$ V nm$^{-1}$ from hole pockets near $M_s$ points (red), hole pockets near $K_s$ points (orange) and electron pockets near $\Gamma_s$ points (green) (**c**). Fermi surfaces at the filling factor $n = 7\frac{2}{3}n_0$ at $D = -0.4$ V nm$^{-1}$ (**b**) and $D = 0$ V nm$^{-1}$ (**d**). The blue (red) line represents the pockets from the **K** (**K′**) valley. The solid black line represents the moiré Brillouin zone.

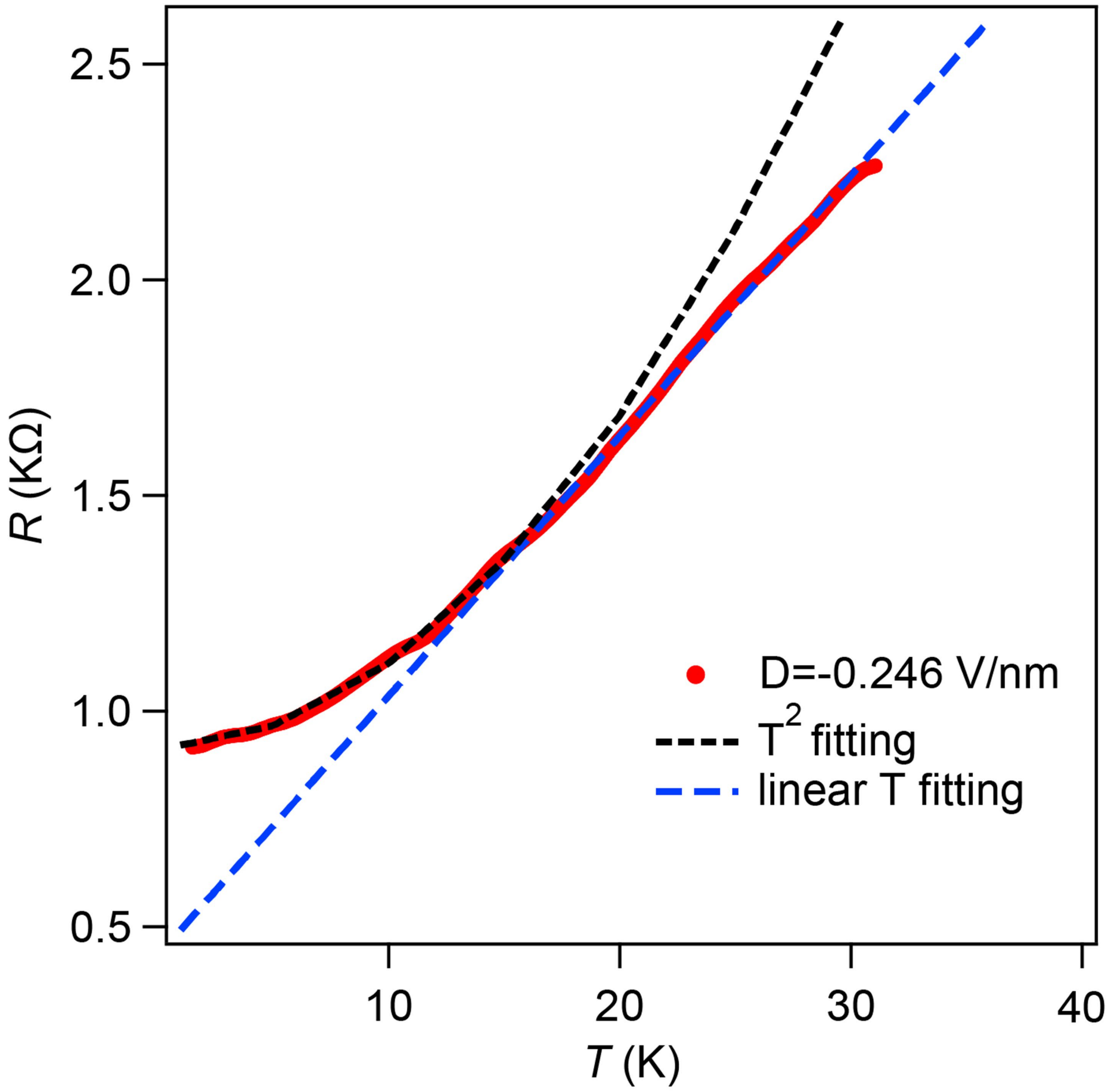

**Extended Data Fig. 7 | Determination of $T_n$.** The $R$–$T$ curve at $n = 7\frac{2}{3}n_0$ and $D = -0.21$ V nm$^{-1}$ is fit well ($r^2 = 0.98$) by a $T^2$ form up to a temperature of 14.7 K and is fit well ($r^2 = 0.996$) by a $T$-linear form.